\documentclass[fleqn,usenatbib,useAMS]{mnras}

\usepackage{graphicx}	
\usepackage{amsmath}	
\usepackage{amssymb}	
\usepackage{multicol}        
\usepackage{bm}		
\usepackage{pdflscape}	

\usepackage[T1]{fontenc}
\usepackage{ae,aecompl}

\usepackage{newtxtext,newtxmath}

\title[Broadened profiles of Diffuse Interstellar Bands]{Broadened profiles of Diffuse Interstellar Bands}

\author[Kre{\l}owski et al.]{
J. Kre{\l}owski,$^{1}$\thanks{jacek@umk.pl, ORCHID:0000-0003-0162-9487}
G.A.~Galazutdinov,$^{2,3}$\thanks{runizag@gmail.com, ORCHID:0000-0003-1188-6487}
P.~Gnaci{\'n}ski,$^{4}$\thanks{piotr.gnacinski@ug.edu.pl, ORCHID:0000-0003-0272-7791}
R.~Hakalla,$^{1}$\thanks{hakalla@ur.edu.pl, ORCHID:0000-0002-3165-8767}
W.~Szajna,$^{1}$\thanks{szajna@ur.edu.pl, ORCHID:0000-0003-4543-1333}
and R.~Siebenmorgen$^{5}$\thanks{Ralf.Siebenmorgen@eso.org, ORCHID:0000-0002-9788-672X}
\\
$^{1}$Materials Spectroscopy Laboratory, University of Rzesz{\'o}w, Pigonia 1 Street, 35-310, Rzesz{\'o}w, Poland\\
$^{2}$Federal State Budget Scientific Institution Crimean Astrophysical Observatory of RAS, Nauchny 298409, Crimea\\
$^{3}$Special Astrophysical Observatory of the Russian AS, Nizhnij Arkhyz 369167, Russia\\
$^{4}$Institute of Theoretical Physics and Astrophysics, Faculty of Mathematics, Physics and Informatics, University of Gda\'nsk, 80-308 Gda\'nsk, Poland\\
$^{5}$European Southern Observatory, Karl-Schwarzschild-Str. 2, 85748 Garching, Germany
}

\date{Accepted XXX. Received YYY; in original form ZZZ}

\pubyear{2021}

\begin{document}
\label{firstpage}
\pagerange{\pageref{firstpage}--\pageref{lastpage}}
\maketitle

\begin{abstract}
The paper describes profile broadening and  peak wavelength variation of diffuse interstellar bands (DIBs) measured for 46 lines of
sight, probably caused by physical properties of intervening clouds. Full width at half maximum of four studied diffuse bands
(5780, 5797, 6196 and 6614 \AA) demonstrate strong variability sometimes doubling the features' width. Despite the high magnitude
of the effect, our current analysis is restricted to the strongest diffuse bands because the weaker ones require a much higher S/N
ratio. The profile broadening in the studied DIBs moves the profile's centers towards longer wavelengths, probably due to the
excitation of higher levels of the P branch of the unknown molecular carrier. Moreover, diffuse bands are broader in clouds
with abundantly populated vibrationally excited states of the hydrogen molecules, i.e. DIB's broadening correlates with the
rotational temperature estimated on H$_2$ $\nu$=2 vibrational level. However, objects demonstrating extremely broadened profiles
of DIBs are scarce. The extreme peculiarity of DIBs' profiles was detected in Herschel~36. Here we
show the gradual growths of the widths of diffuse bands, confirmed in spectra from different instruments.
\end{abstract}


\section{Introduction}

The paper concerns physics and chemistry of the interstellar medium
(ISM) \textemdash which consists of several components: {\it gas
atoms, molecules, dust grains}. Overall it addresses a
scientifically very challenging issue of identifying (very likely
molecular) carriers of the puzzling spectral features known, since
nearly 100 years, as diffuse interstellar bands (DIBs). See \cite{JK18} for a
recent review.  The only almost confirmed carrier of a few near infrared diffuse bands
is fullerene C$_{60}^+$ (Campbell et al. 2015) but the variability of the intensity ratio of
features, assigned to this molecule, is still waiting for an explanation (Galazutdinov  et al. (2017, 2020);
Galazutdinov \& Kre{\l}owski 2017). 

DIBs are believed to be carried by complex chemical species, including
prebiotic ones, synthesized and preserved in the hostile
conditions of translucent interstellar clouds. The idea of their
molecular origin is supported by presence of substructures inside
DIB profiles, discovered by \cite{Sarre1995} and by \cite{Kerr1998};
for weak diffuse bands see \cite{GLK08}; for broad diffuse bands see Galazutdinov et al. (2020).
A vast majority of DIBs are narrow, though much broader than atomic lines,
shallow features and thus an analysis of their profile shapes
requires a very high resolution and S/N ratio. High precision DIB
profiles may play a decisive role in the carrier's identification.
However, the studied line of sight must demonstrate the lack of Doppler
splitting in lines of interstellar atoms and simple radicals to secure
the intrinsic profile of a diffuse band.
There are more than 200 molecular species discovered (mostly in radio) due to rotational
transitions; a vast majority of them being carbon--bearing (organic) ones.
The list of DIBs, seen in absorption, reaches now $\sim$560 entries
\citep{Fan2019}.

Experiments show that spectra of suspected molecules contain usually one
strong and several weak features. It is very challenging
to relate weak DIBs to the strong ones and thus to divide them into
``families'', likely sharing the same carrier. Their profiles can vary
with temperature (kinetic and rotational) in the way characteristic to
a given species. Changes of profiles may lead to different
central wavelengths. It is important to relate DIB profiles (their
widths, substructure patterns and intensities) to relative abundances of CH,
CH$^+$, CN and rotational temperatures of the C$_2$ and C$_3$
molecular species; perhaps also to that of the CN.
Diffuse bands carriers are not known, thus any relation with other measurable
 interstellar parameters/features may give a clue about the origin of these features.

The results published during the last two decades (e.g.
\cite{KG99},  \cite{G06}, \cite{GLK08},
and \cite{KGMMC15}) demonstrate red- and blue-shifts of some DIBs in
relation to atomic lines (i.e. not caused by the Doppler effect) as well as changes of their profiles,
which vary from one to another specific environment. This was
already supported by finding a relation between rotational
temperatures of simplest carbon chains and profile shape of the 6196 DIB \citep{K2009}.

In \cite{K2010} authors reported that width and shape of 6196 and 5797 \AA\ DIB profiles
depend on the gas kinetic and rotational temperatures of C$_2$ molecule; the profiles are broader because
of the higher values of temperature. However, this phenomena is not common for all DIBs: e.g. 4964 and 5850 \AA\ features do not exhibit detectable broadening.

In a vast majority of cases one can observe reddened stars through
several clouds threaded on their sightlines. This leads to
ill--defined averages which closely resemble each other. On the
other hand, if a star is seen through a single cloud, its interstellar spectrum looks ``peculiar''. Such spectra are especially
interesting as they are apparently formed in a specific volume, i.e. in reasonably homogeneous physical conditions.
This concerns e.g. the above mentioned wavelength shifts. The latter are
especially evident in the unique object \textemdash  Herschel
36 \citep{Dahlstrom2013,OWJ13}. DIBs in its spectrum exhibit extremely
uncommon profiles, in particular showing extended red wings. Identified interstellar lines, as shown by Dahlstrom {\it et~al.},
are sharp and free of Doppler splitting. Several DIB profiles,
observed along the specific sightline (to Herschel 36), show unusually broadened
profiles (e.g. the major 5780 DIB).

The presence of substructures in profiles of diffuse bands is the
common property which can be useful for identifying their origin by
comparing astrophysical data with laboratory gas phase spectra, as for example
the  carbon chains or Polycyclic Aromatic Hydrocarbons
\textemdash  PAHs \citep{Salama1999,Motylewski2000}.

It is important to check the behaviour of identified and
unidentified interstellar spectral features, especially if they are
observed in single clouds, where the physical conditions are as
homogeneous as possible. Such sightlines should be carefully selected and confirmed by
atomic/molecular lines, seen in high resolution spectra, as free of Doppler splitting.

\section{Observational material}

The spectra we selected do not show evident Doppler split in
atomic/molecular lines of interstellar origin observed with
instruments enlisted below. Usually we use the K{\sc i} 7699 \AA\
and CH 4300 \AA\ lines. Intensity of these lines generally
demonstrate quite good correlation with those of DIBs. However, we
have to emphasize that with increasing spectral resolution one can
resolve additional Doppler components in atomic and/or molecular
lines; in diffuse bands it becomes very difficult and in case of
extinction or polarization (caused by dust grains) \textemdash
impossible. Nevertheless, variability of the full width at the half
maximum (FWHM) of diffuse bands reported in this study is not
caused by the presence of several clouds with different radial
velocities.

The observational data have been collected using several high resolution, echelle spectrographs:
\begin{itemize}
\item
     UVES (Ultraviolet and Visual Echelle Spectrograph) fed by the 8m Kueyen VLT mirror
     \citep{Dekal00}. The spectral resolution is up to R=80,000 in the blue range and R=110,000 in the red one.
     The telescope size allows to get high S/N ratio spectra of even pretty faint stars. The selected spectra have been collected
     in the frame of the Large Program EDIBLES (ESO Diffuse Interstellar Bands Large Exploration Survey) and our previous study
     (Siebenmorgen et al. 2020) where we provide online access to the analyzed UVES data.\footnote{https://vizier.u-strasbg.fr/viz-bin/VizieR?-source=J/A+A/641/A35}
\item
     Some southern objects were studied with the aid of Feros - fiber fed echelle
     spectrograph (Kaufer et al. 1999) at the ESO's La Silla observatory in Chile. Feros
     provides the resolving power R$\equiv$$\lambda$/$\Delta$$\lambda$
     of 48,000 and allows to get the whole available spectral range
     ($\sim$3700 -- 9200~\AA, divided into 37 orders) recorded in a
     single exposure.
\item
    ESPaDOnS spectrograph (Echelle SpectroPolarimetric Device for the
    Observation of Stars)\footnote{https://www.cfht.hawaii.edu/Instruments/Spe\-ctroscopy/Espadons/} is the bench-mounted high-resolution echelle
    spectrograph/spec\-tro\-pola\-ri\-meter) attached to the 3.58~m
    Canada-France-Hawaii telescope (CFHT) at Mauna Kea (Hawaii, USA). It is
    designed to obtain a complete optical spectrum in the range from
    3,700 to 10,050~~\AA. The whole spectrum is divided into 40 \'{e}chelle
    orders. The resolving power is about 68,000.
\item
  The CFHT coud\'{e} spectrograph (Gecko)\footnote{https://www.cfht.hawaii.edu/Instruments/Spectroscopy/Gecko},
  an echelle spectrograph optimized for use with a single spectral order from the 316 groove/mm echellette mosaic
  and providing resolving power $\lambda$/$\Delta\lambda$ up to 120,000. Order sorting is achieved with interference
  filters or by one of three variable grisms. An image slicer is used to optimize the throughput of the instrument.
\item
 HARPS spectrograph \citep{May03}, fed by the 3.6m ESO telescope at La Silla
 observatory with resolving power R=115,000.
\item
 MIKE spectrograph  (Bernstein et al. 2003)  fed by the Magellan/Clay telescope.
 The spectral resolution with a 0.35$\times$5 arcs slit ranges from
 $\sim$56,000 on the blue side (3600-5000~~\AA) to $\sim$77,000 on the red side (4800-9400~~\AA).
\item
    The Bohyunsan Echelle Spectrograph (BOES) of the Korean National Observatory \citep{kimetal2007} is installed at the 1.8m telescope of the Bohyunsan Observatory in Korea. The spectrograph has three
    observational modes allowing resolving powers of 30,000, 45,000, and
    90,000. In any mode, the spectrograph covers the whole spectral range of
    $\sim$3500 to $\sim$10,000~\AA, divided into 75 -- 76 spectral
    orders.
\item
    The Sandiford Cassegrain Echelle Spectrometer (McCarthy et al. 1993) fed by 2.1-meter Otto Struve Telescope at McDonald Observatory (Texas, USA). The instrument has a resolving power R = 60,000 for two CCD pixels
    and provides continuous wavelength coverage below  8000~\AA.
\end{itemize}

\begin{table*}
\caption{Basic parameters for selected stars and
  the central wavelength (the middle point of FWHM) and the FWHM (km/s) of 5780, 5797, 6614 and 6196 diffuse bands.
  Distances are given as reciprocals of the Gaia DR3 parallaxes.
  Bold faced targets are marked by red circles in Fig. 2. For targets with the lack of Gaia distances (marked with "Ca"), the latter were
  measured with the aid of Ca{\sc ii}-method (Megier et al. 2005, 2009).
  Objects with blue-shifted DIBs are given separately in the bottom part of the table.
  Ending letters in the star names label the origin of analyzed spectra: b - BOES, h - HARPS, e - ESPADONS, f- FEROS, m - MIKE, s - McDonald, u - UVES.
}
\centering{
\begin{tabular}{rlccclrlrlrlr}
\hline
star   & Sp/L  &  V   &  B-V  & D(Gaia)& $\lambda_{cen}$ & FWHM   &$\lambda_{cen}$ & FWHM   &$\lambda_{cen}$ & FWHM   &$\lambda_{cen}$ & FWHM    \\
       &       & [mag]& [mag] & [pc]   &   [\AA]         & [km/s] &[\AA]           & [km/s] &[\AA]           & [km/s] &[\AA]           & [km/s]  \\
\hline
{\bf 23180e}  & B1III  & 3.86&0.02&330  & 5780.41 &103 & 5797.01 & 29 & 6613.58 & 41 & 6195.97 & 19 \\
22951u  & B0.5V & 4.97 & -0.06 &  370   & 5780.39 &102 & 5797.02 & 35 & 6613.55 & 41 & 6195.95 & 18 \\
24263u  & B3.5V & 5.77 &  0.03 &  225   & 5780.43 &103 & 5797.05 & 35 & 6613.57 & 42 & 6195.99 & 18 \\
24398b  & B1Ib  & 2.85 &  0.12 &  260   & 5780.38 &103 & 5797.02 & 36 & 6613.56 & 42 & 6195.92 & 19 \\
27778m  & B3V   & 6.34 &  0.16 &  210   & 5780.43 &105 & 5797.01 & 34 & 6613.57 & 43 & 6195.94 & 20 \\
30492u  & B9.5V & 9.02 &  0.28 &  435   & 5780.44 &104 & 5797.01 & 34 & 6613.57 & 42 & 6195.97 & 19 \\
36695u  & B9IV  & 5.34 & -0.18 &  440   & 5780.44 &102 & 5797.11 & 53 & 6613.63 & 52 & 6195.92 & 20 \\
{\bf 36982u}& B1.5V &8.46&0.11 &  410   & 5780.84 &127 & 5797.2  & 60 & 6613.77 & 57 & 6195.98 & 27 \\
37020e  & B0V   & 6.73 &  0.02 &  380   & 5780.84 &120 &  	    &    &		   &    & 6195.97 & 17 \\
37021e  & B1V   & 7.96 &  0.24 &  375   & 5781.10 &134 & 	    &    &         &    &         &    \\
37022m  & O8V   & 5.13 &  0.02 &  400   & 5780.88 &118 & 	    &    &         &    & 6196.03 & 24 \\
37023e  & B1.5V & 6.70 &  0.09 &  440   & 5780.89 &127 & 	    &    &         &    & 6196.00 & 19 \\
37041u  & O8V   & 6.30 & -0.09 &  335   & 5780.74 &120 & 	    &    &         &    &         &    \\
37042s  & O9.2Iab& 6.38& -0.09 &  420   & 5780.65 &116 & 	    &    &         &    &         &    \\
37061e  & O9V   & 6.83 &  0.22 &  415   & 5780.92 &124 & 5797.16 & 50 & 6613.96 & 74 & 6196.04 & 28 \\
37128u  & B0Ia  & 1.69 & -0.18 &{\bf 330Ca}&5780.77 &123 &       &    & 6613.87 & 63 &         &    \\
37130u  & B9V   &10.10 &  0.05 &  395   & 5780.92 &130 & 5797.12 & 46 & 6613.83 & 64 & 6195.94 & 26 \\
37903b  & B3V   & 7.83 &  0.11 &  400   & 5780.92 &130 & 5797.12 & 46 & 6613.83 & 64 & 6195.94 & 26 \\
66194u  & B3V   & 5.81 & -0.09 &  405   & 5780.44 &109 & 5797.05 & 42 & 6613.50 & 41 & 6195.93 & 20 \\
110432u & B0.5IV& 5.31 &  0.27 &  440   & 5780.45 &102 & 5797.02 & 34 & 6613.58 & 41 & 6195.97 & 18 \\
110715u & B9V   & 8.65 &  0.38 &  525   & 5780.34 &103 & 5796.99 & 36 & 6613.57 & 43 & 6195.95 & 17 \\
133518e & B2IV  & 6.39 & -0.10 &  610   & 5780.45 &102 & 5797.06 & 36 & 6613.61 & 46 & 6195.96 & 18 \\
143275b,s & B0.3IV&2.32&-0.12&{\bf 210Ca} & 5780.44 & 97 & 5796.99 & 34 & 6613.65 & 52 & 6195.94 & 23 \\
144217h & B1V   &2.62&-0.07&{\bf 235Ca} & 5780.43 &103 & 5797.00 & 33 & 6613.52 & 40 & 6195.94 & 20 \\
144470b & B1V   & 3.97 & -0.05 &  140   & 5780.46 &102 & 5797.04 & 34 & 6613.59 & 45 & 6195.97 & 21 \\
145502m & B2V   & 4.00 &  0.05 &  140   & 5780.43 &104 & 5797.03 & 36 & 6613.57 & 43 & 6195.95 & 21 \\
{\bf 147165h}& B1III&2.89&0.13 &  100:  & 5780.50 &106 & 5797.07 & 39 & 6613.58 & 44 & 6195.95 & 22 \\
147888u & B3V   & 6.74 &  0.31 &  125   & 5780.55 &108 & 5797.06 & 37 & 6613.71 & 53 & 6195.97 & 24 \\
{\bf 147889h}&B2.5V&7.90&0.83&    135   & 5780.64 &113 & 5797.07 & 41 & 6613.59 & 46 & 6195.95 & 24 \\
147932u & B5V   & 7.27 &  0.32 &  125   & 5780.59 &116 & 5797.09 & 42 & 6613.67 & 53 & 6195.94 & 22 \\
147933e & B1V   & 5.05 &  0.17 &  140   & 5780.55 &109 & 5797.08 & 42 & 6613.64 & 51 & 6195.93 & 25 \\
148184e & B2V   & 4.43 &  0.28 &  150   & 5780.41 & 95 & 5797.08 & 43 & 6613.56 & 45 & 6195.94 & 25 \\
148579u & B8V   & 7.32 &  0.26 &  140   & 5780.57 &111 & 5797.11 & 45 & 6613.73 & 55 & 6195.96 & 25 \\
148605u & B3V   & 4.79 & -0.07 &  125   & 5780.60 &108 & 5797.10 & 40 & 6613.70 & 59 & 6195.94 & 21 \\
149757e & O9.2IV& 2.56 &  0.02 &  135   & 5780.40 &105 & 5797.01 & 36 & 6613.54 & 42 & 6195.95 & 24 \\
154445u & B1IV  & 5.61 &  0.12 &  250   & 5780.40 &100 & 5797.03 & 35 & 6613.55 & 40 & 6195.96 & 17 \\
163800h &O7.5III& 7.00 &  0.27 & 1310   & 5780.41 &104 & 5797.00 & 33 & 6613.53 & 41 & 6195.94 & 18 \\
164906u & B1IV  & 7.38 &  0.23 & 1345   & 5780.44 &105 & 5797.04 & 33 & 6613.58 & 42 & 6195.97 & 17 \\
170634u & B6.5V & 9.85 &  0.58 &  440   & 5780.41 &101 & 5797.03 & 43 & 6613.53 & 44 & 6195.90 & 20 \\
170740u & B2V   & 5.72 &  0.24 &  225   & 5780.45 &102 & 5797.02 & 32 & 6613.57 & 41 & 6195.97 & 18 \\
179406h & B2.5II& 5.34 &  0.13 &  290   & 5780.40 &101 & 5797.02 & 31 & 6613.57 & 39 & 6195.96 & 17 \\
184915u &B0.5III& 4.96 & -0.03 &  505   & 5780.44 &102 & 5797.07 & 36 & 6613.57 & 40 & 6195.93 & 16 \\
185418u & B1IV  & 7.49 &  0.15 &  705   & 5780.42 &101 & 5797.04 & 34 & 6613.57 & 41 & 6195.97 & 17 \\
185859b & B0Iab & 6.59 &  0.37 & 1035   & 5780.42 &105 & 5797.00 & 35 & 6613.54 & 43 & 6195.92 & 18 \\
203532m & B3.5IV& 6.38 &  0.13 &  290   & 5780.35 & 99 & 5797.01 & 36 & 6613.56 & 44 & 6195.92 & 17 \\
210121u & B3.5V & 7.68 &  0.16 &  335   & 5780.50 & 97 & 5797.05 & 40 & 6613.61 & 50 & 6195.96 & 22 \\
\hline
34078b  & O9.5V & 5.96 &  0.22 &   389  & 5780.45 & 109& 5797.04 & 51 & 6613.52 & 51 & 6195.83 & 23 \\
152218f & O9V   & 7.57 &  0.21 &  1591  & 5780.36 & 105& 5797.00 & 48 & 6613.47 & 45 & 6195.81 & 21 \\
152076f &B0/1III &8.90 &  0.20 &  1674  & 5780.32 & 103& 5796.97 & 39 & 6613.47 & 46 & 6195.82 & 20 \\
152200f &O9.7IV(n)&8.39 &0.10  &  1447  & 5780.36 & 106& 5797.00 & 47 & 6613.50 & 47 & 6195.84 & 22 \\
152233h &O6II(f)& 6.59 & 0.13  &  1436  & 5780.30 & 104& 5796.97 & 41 & 6613.43 & 45 & 6195.81 & 20 \\
152234f &B0.5Ia & 5.45 & 0.19  &{\bf 1823 Ca}& 5780.32 & 105& 5796.96 & 45 & 6613.47 & 46 & 6195.82 & 23 \\
152246f & O9IV  & 7.29 &  0.16 &  1886  & 5780.33 & 104& 5796.99 & 43 & 6613.48 & 45 & 6195.83 & 21 \\
152314f & O9IV  & 7.75 & 0.22  &  1392  & 5780.33 & 108& 5796.96 & 47 & 6613.48 & 47 & 6195.82 & 21 \\
\hline
\end{tabular}
}
\label{targets}
\end{table*}

\begin{table}
\centering{
\caption{Amplitude of variation (minimum and maximum) of the FWHM's center
and the FWHM for four measured diffuse bands.
}
\begin{tabular}{cccc}
\hline
   \multicolumn{2}{c}{$\lambda_{cen}$(\AA)}
                      &   \multicolumn{2}{c}{FWHM(km/s)}   \\
  min    & max        &  min    & max \\
\hline
 5780.34  & 5781.10   &   95    & 134 \\
 5796.99  & 5797.20   &   29    &  60 \\
 6613.52  & 6613.96   &   39    &  74 \\
 6195.90  & 6196.04   &   16    &  28 \\
\hline
\end{tabular}
\label{amplitude}
}
\end{table}

We processed raw data and made measurements in the reduced spectra
  with our interactive analysis software DECH\footnote{http://www.gazinur.com/DECH-software.html}.
  For the DECH data reduction, we averaged bias images for subsequent correction of
  all other images. The scattered light was determined as a complex
  shaped two-dimensional surface function, which is individually
  calculated for each stellar and flat-field frame by a 2D cubic-spline
  approximation over areas of minima between the spectral orders.  Then,
  the pixel-to-pixel variations across the CCD were corrected by
  dividing all stellar frames by the averaged and normalized flat-field
  frame. One-dimensional stellar spectra were extracted by simple
  summation in the cross-dispersion direction along the width of each
  spectral order. The extracted spectra of the same object, observed in
  the same night, were averaged to achieve the highest S/N ratio.
  Fiducial continuum normalization was based on a cubic spline
  interpolation over the interactively selected anchor points.

\section{Results}

\begin{figure}
\includegraphics[width=9cm]{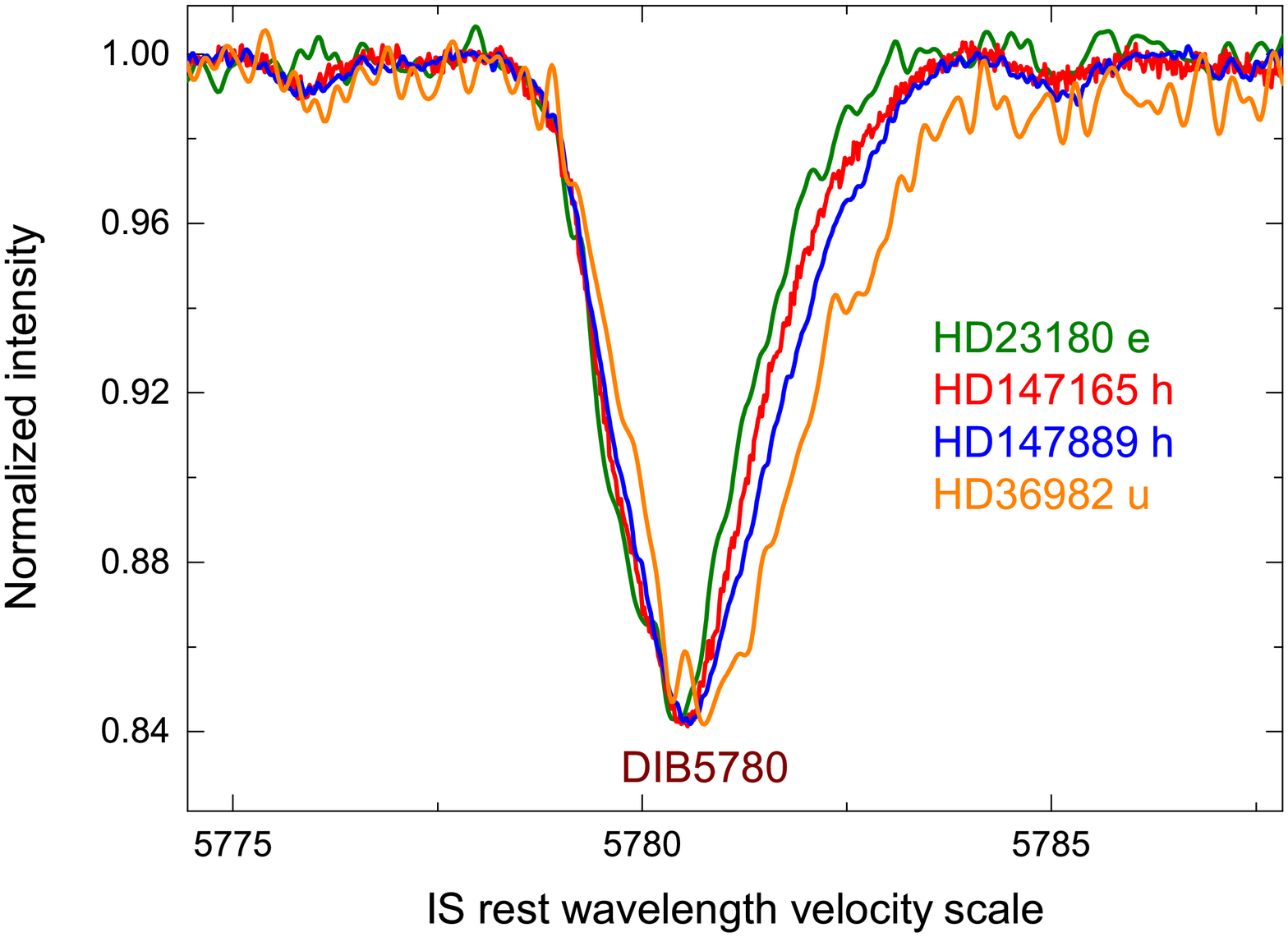}
\caption{The 5780 DIB profiles in four targets normalized to common central depth for clarity.
The profiles are shifted to the rest wavelength velocity frame using the interstellar K{\sc i} line.
The FWHMs of individual profiles evidently differ from object to object; the profiles show also
some wavelength shifts. }
\label{fig1}
\end{figure}

For our analysis we have selected 46 reddened stars with the lack of strong Doppler split in the interstellar K{\sc i} 7700 \AA\ line and/or CH 4300 \AA\ (Table \ref{targets}).
The sample includes the objects from the Orion Trapezium where narrow DIBs are broadened but very weak.
Measurements of FWHM (km/s) of 5780, 5797, 6614 and 6196 diffuse bands and the midpoints of FWHM ($\lambda_{cen}$)
are given in the same Table. All diffuse bands exhibit more or less significant variability of the profile widths (Table 1, 2).
However, our sample is statistically significant thus the inferred relations are important.

The oldest known two diffuse bands 5780 and 5797 have been demonstrated as being of different origin (\cite{KG87}, \cite{KW88}).
Their varying strength ratio allows to divide the interstellar clouds into $\sigma$ and $\zeta$ types.
Some of them are shown in Fig. \ref{fig1}, demonstrating ``peculiar'' behaviour of
the 5780 DIB in certain objects.  The depicted ones represent extreme $\sigma$ or $\zeta$ types, i.e. the 5780/5797 strength ratio is very high or low.
Apparently the same DIB may be narrower or broader (depending on the line of sight); since the depicted targets do not show strong Doppler split in
interstellar K{\sc i} line, this must be caused by physical parameters of the intervening objects.

\begin{figure*}
\centering{
\includegraphics[width=14cm]{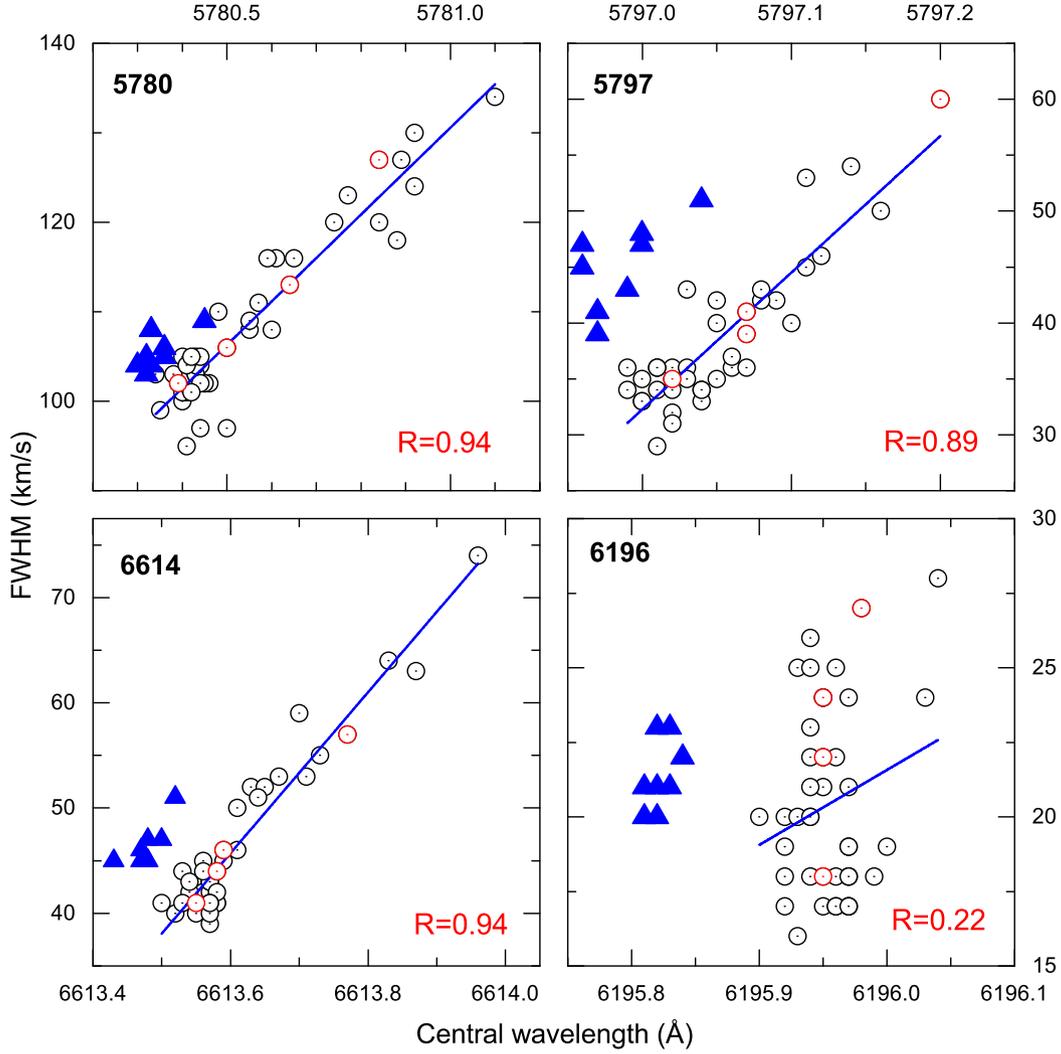}
\caption{
FWHM of diffuse bands versus central wavelength (from the Table
\ref{targets}). The objects, depicted in Figure \ref{fig1}, are
marked by red circles. Note the lack of correlation (R) for the
narrowest known 6196\AA DIB.
Blue triangles correspond to the objects with blue-shifted diffuse bands. These targets were not included to the fit and the estimation of correlation.  }
\label{4dibs}
}
\end{figure*}

Among the observed objects one can see evident positive correlation
between the FWHM and the $\lambda_{cen}$ depicted in Fig. \ref{4dibs}. Three of them increase their width propagating towards
longer wavelengths, probably due to the excitation of higher and
higher levels of the P branch of the unknown molecular carrier. The
same effect is evident for 6196 DIB too (see Fig.4) but its
extended red wing in most of the cases does not reach the 50\%
depth of the feature (see below) and, therefore the midpoint of the
FWHM is almost stationary and resulting formal correlation is low
(R=0.22, see Fig. \ref{4dibs}).

A particular problem is so called blue-shift of some diffuse bands  reported for HD 34078 and several
stars in Sco OB1 association (Galazutdinov et al. 2006, 2008; Kre{\l}owski et al. 2019b).
The blue-shift does not resemble the red-shift effect, i.e. it is not an extension of the profile's blue wing
but is rather a displacement of the entire profile (see e.g. Fig.5 in Kre{\l}owski et al. 2019b).
We suggest that blue-shift may have surprisingly simple explanation.
The "interstellar" wavelength scale based on the strongest component of CH 4300 or KI 7698 line might be slightly
incorrect for these objects. Indeed, all objects demonstrating the blue-shift also exhibit some weak component in the left side of their CH and/or KI profile.
At the moment, there are known very few objects with evident blue-shift of DIB profiles, thus the extension of the  sample is necessary for the decisive
identification of the origin of the blue-shift.

Table \ref{targets} contains data of the central wavelength measured for 5780, 5797, 6614 and 6196 DIBs as well as their FWHM.
This $\lambda_{cen}$ is evidently variable (see Fig. \ref{fig1}).
Moreover, the width of 5797, 5780, and 6614 DIBs increases to the longer wavelength side what is shown in the gradual move of the $\lambda_{cen}$
to the red with increasing the width of diffuse bands.

%

\begin{figure}
\includegraphics[width=9cm]{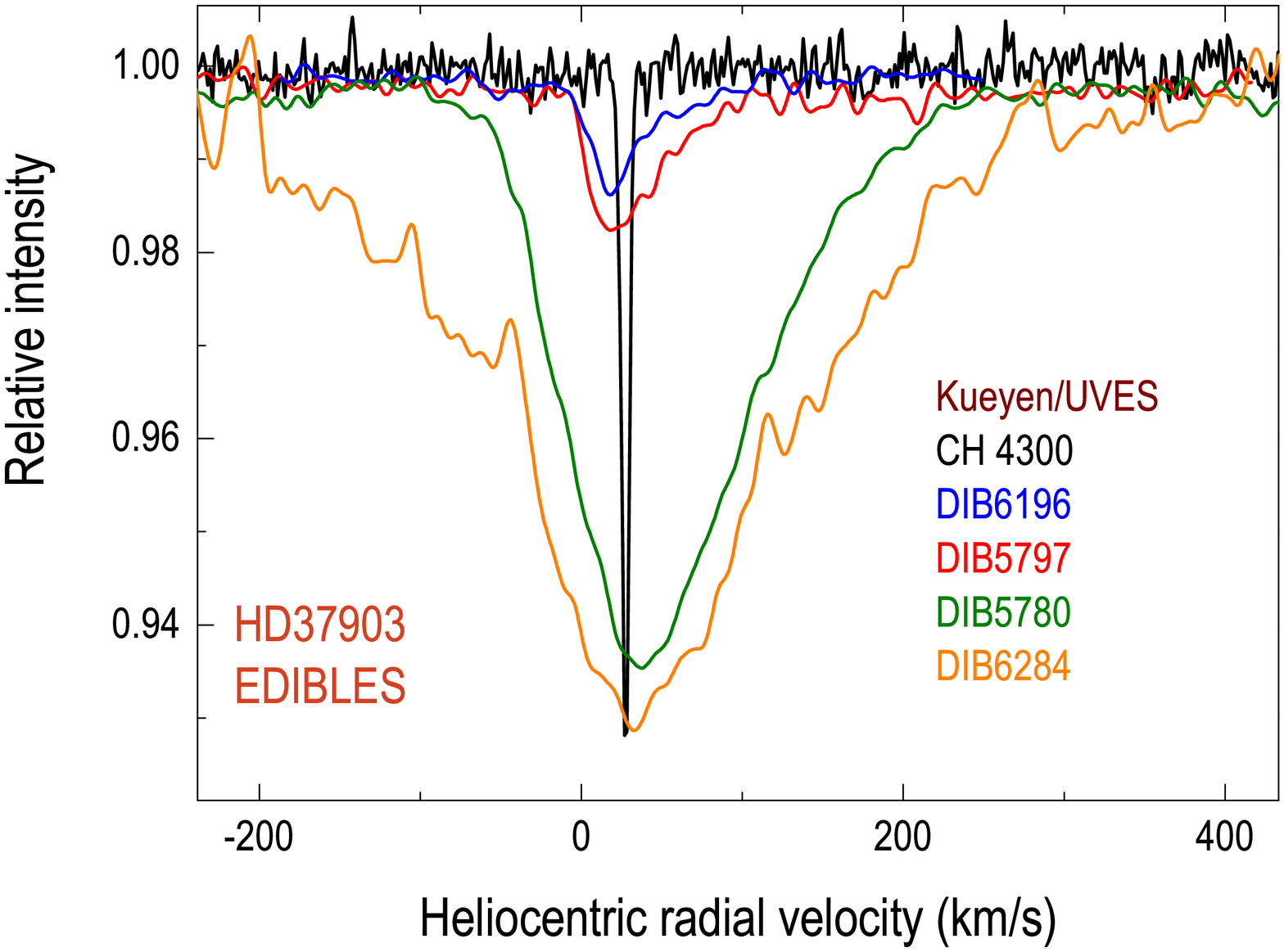}
\caption{Broad and narrow DIBs in heliocentric radial velocity scale in UVES spectrum of HD~37903. The broad ones are
red-shifted by about $\sim$20 km/s in relation to the
position of CH line while the narrow (weak) DIBs remain at the
normal central position or even are slightly blue-shifted.}
\label{shift}
\end{figure}

An additional effect, moving the central wavelengths of diffuse bands to the right (red-shift), is observed in some targets.
An example is shown in Fig. \ref{shift}. A vast majority of DIBs are
very weak in the spectra of stars like HD~37903 with low or moderate reddening magnitude. Only the relatively broad DIBs,
such as 5780 and 6284, remain strong; and they are also red shifted. The red-shift phenomenon is observable only in these broad DIBs.
The central wavelengths of the weak narrow diffuse bands seem even to be slightly blue-shifted.
Figure \ref{shift} demonstrates the DIB positions in the heliocentric radial velocity scale, compared to the interstellar CH (4300.3~\AA) interstellar line.

\begin{figure}
\includegraphics[width=9cm]{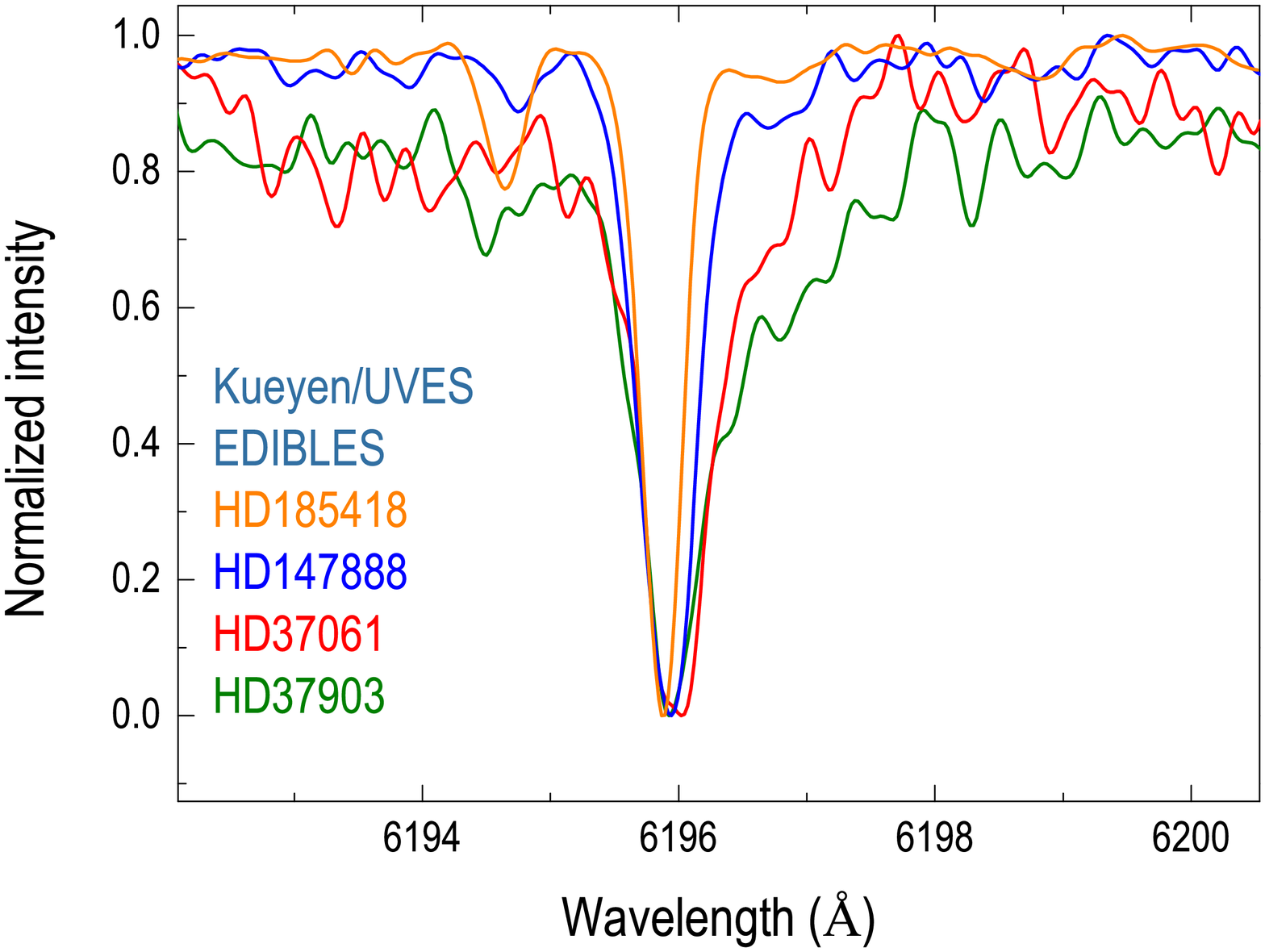}
\caption{
The 6196 DIB profiles normalized to common central depth. HD~147888 and HD~185418 are
comparison objects though the former also shows some DIB broadening.
Note the presence of an extended red wing in HD~37061 and HD~37903.}
\label{6196}
\end{figure}

\begin{figure}
\includegraphics[width=9cm]{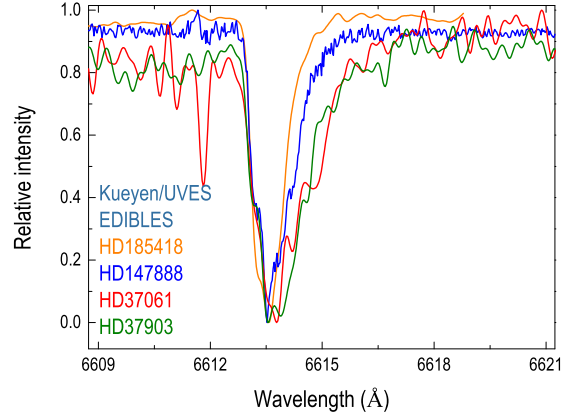}
\caption{The 6614 DIB profiles normalized to common central depth. HD~147888 and HD~185418 are
comparison objects though also showing some DIB broadening.}
\label{6614}
\end{figure}

\begin{figure}
    \centering
		\includegraphics[width=9cm]{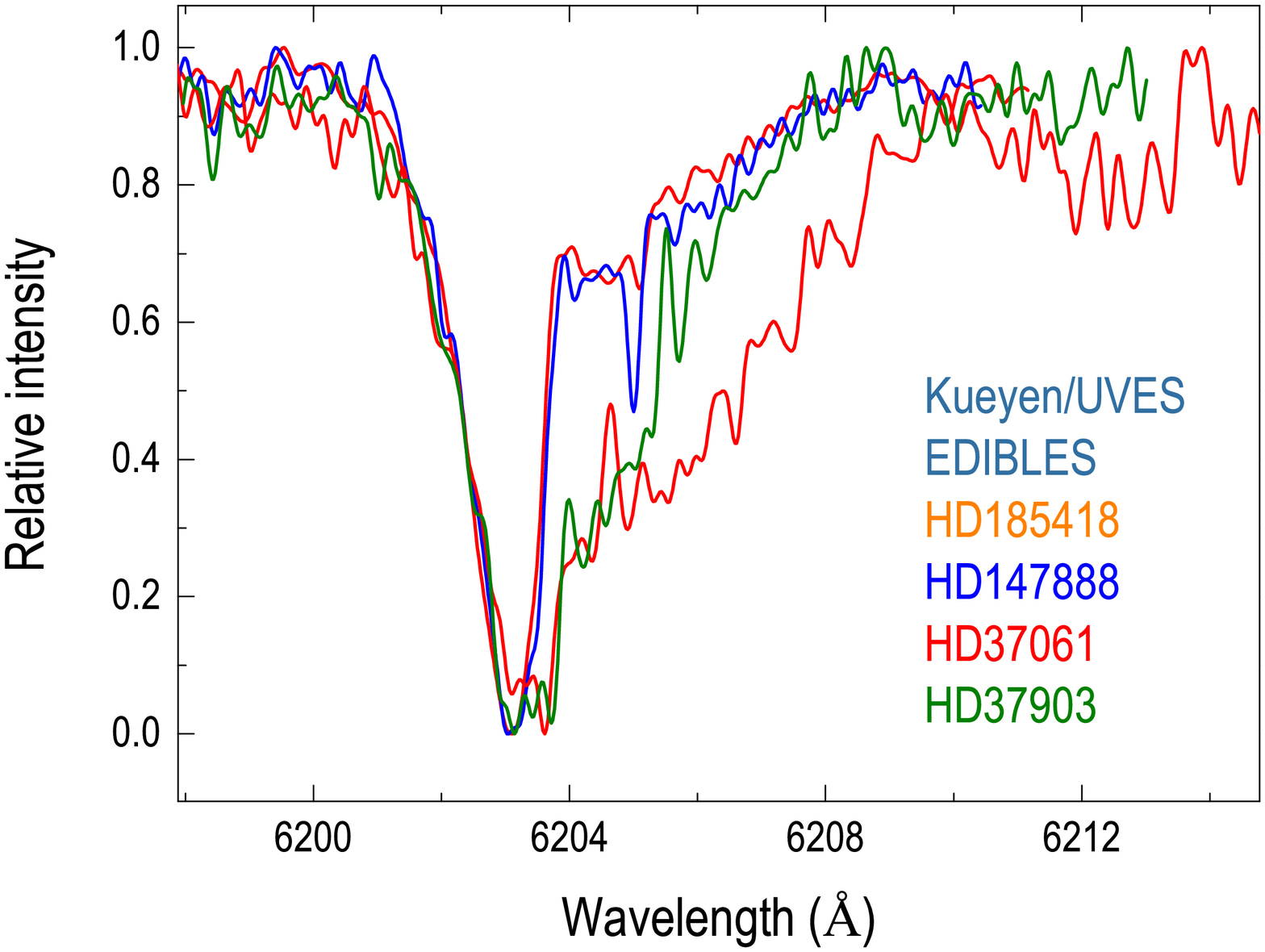}
    \caption{The 6203 DIB profiles normalized to common central depth. HD~147888 and HD~185418 are
comparison stars though also showing some DIB broadening. Perhaps this is a blend of two features, centered at
6203 and 6205~~\AA. }
    \label{6203}
 \end{figure}

The rest of narrow diffuse bands are most likely broadened though not shifted. However,
narrow DIBs in the extreme $\sigma$ type objects are so weak, that any conclusions concerning
them must be considered as hardly reliable. A much higher S/N ratio would be of great importance.

\section{Molecular hydrogen}

 \begin{table}
  \caption{Temperatures of molecular hydrogen. The T$^{para}_{\nu=2}$ is the
	rotational temperature of para--H$_2$ on the vibrational level $\nu$=2.
	For HD 37061 and HD 147888 the reference is the source of H$_2$ column densities used to
	calculate the temperature. }
  \begin{tabular}{lrrl}
	\hline
    star                & T$_{01}$ [K] & T$^{para}_{\nu=2}$ [K] & reference \\
  \hline  	
		HD 27778	    & 55$\pm$5    & ---          & \cite{Jensen} \\
		HD 37061	    & ---         & 2088$\pm$231 & \cite{Gnacinski2009} \\
		HD 37903	    & 67$\pm$8    & 1779$\pm$210 & \cite{Gnacinski2011} \\
		HD 147888       & 42$\pm$3    & 791$\pm$128  & \cite{Gnacinski2013} \\
		HD 185418	    & 100$\pm$9   & ---          & \cite{Jensen} \\
  \hline
	\end{tabular}
	\label{temp}
 \end{table}

Gnaci\'{n}ski et al. (2016) analyzed correlation between temperature of H$_2$ and equivalent widths of
some diffuse bands normalized by the total amount of interstellar hydrogen EW(DIB)/N(H).
It was shown that in most cases the EW(DIB)/N(H) and temperatures are anti-correlated. Here we are focused on the comparison
of temperature of molecular hydrogen with the  peak wavelength variation and broadening of diffuse bands.

 The H$_2$ absorption lines from vibrationally
excited levels can be seen in some HST (Hubble Space Telescope) ultraviolet spectra.
We used the temperatures of para \textemdash H$_2$ on the $\nu$=2 vibrational level,
calculated using linearization of the Boltzmann equation and linear regression
as described by \cite{Gnacinski2011}. These temperatures are presented
in Table \ref{temp}.

The cloud heliocentric radial Doppler velocities of the H$_2$ lines agree with  those of atomic K{\sc i} lines seen in the UVES spectra.
Usually the Doppler velocities of simple radicals (CH, CN) coincide with those of K{\sc i};
thus we suggest that DIB carriers are occupying same volumes as the above mentioned species.

It is interesting that difference in the widths of diffuse bands are not caused by the gas kinetic temperature, which is correlated with the hydrogen T$_{01}$ temperature.
The latter originally defined by Spitzer, Dressler, \& Upson (1964) describes the equilibrium between
ortho and para-H$_2$ and is calculated from the equation:

\begin{equation}
 \frac{N_1}{N_0}=9e^{\frac{-(E_1-E_0)}{kT_{01}}}
\end{equation}

where N$_1$ and N$_0$ are the number of molecules per cm$^2$ in the levels of angular momentum 1 and 0 respectively, E$_1$ and E$_0$ are the excitation energy of the levels J = 1 and J=0 respectively.

The comparison star HD~185418 that shows relatively narrow diffuse bands has the T$_{01}$=100~K which is much higher than the T$_{01}$ temperatures towards HD~147888 or HD~37903 (42~K and 67~K, see Tab. \ref{temp}).
The temperature T$_{01}$ observed towards HD~27778 is only 55~K. Despite the lack of direct correlation between the DIB's FWHM  and the gas kinetic temperature
in the 6196 DIB (Fig. \ref{6196}), the broadening of line profile is seen mainly for stars with highest T$^{para}_{\nu=2}$
temperatures: HD~37061 and HD~37903. In both of the above objects the 6196 DIB profile shows the extended red wing.

  Two comparison stars HD 147888 and HD 185418 	show nearly unbroadened profiles of the diffuse interstellar bands. Both stars have
UVES and HST high dispersion spectra. 	We have used HST STIS (Space Telescope Imaging Spectrograph)
spectra {\it o59s01010} and {\it o59s01020} to check the
presence of H$_2$ lines from vibrationally excited levels in the
direction towards HD 147888. For another comparison star, HD 185418,
we used the spectra {\it ob2609010}, {\it ob2609020}, {\it o5c01q010} and {\it obkr1q010}.
In the direction of both stars no H$_2$ absorption lines from
vibrationally excited levels were detected. The H$_2$ is present
towards both comparison stars on the ground vibrational level. The column densities of the $\nu$=0 H$_2$ rotational levels towards these stars were adopted from \cite{Jensen}.

FWHM of the 5780 DIB in both  comparison stars are about $\sim$100-110 km/s, i.e. smaller than that
in ``peculiar'' targets (Tables \ref{targets}, \ref{amplitude}). Consequently, the broadening of the 6614 and 6203 DIBs (Figs. \ref{6614}
and \ref{6203}) is larger for HD~37061 than for HD~37903. The cloud towards HD 37061 has the highest T$^{para}_{\nu=2}$ temperature in our sample.
The direction towards HD 147888 shows no broadening of the 6203 DIB, and
only slight broadening of 6614 DIB. This is consistent with lower T$^{para}_{\nu=2}$
temperature towards this star.

\section{Instrumental comparison}

\begin{figure}
\includegraphics[width=9cm]{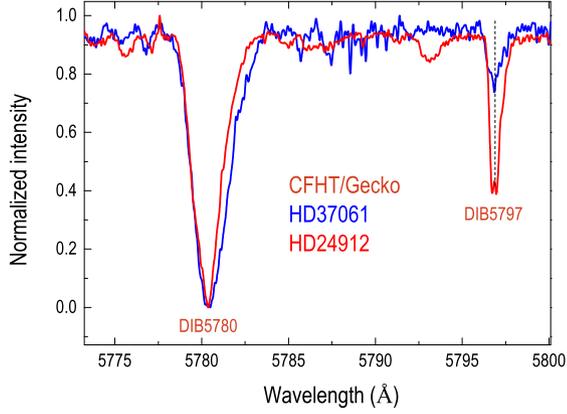}
\caption{The major diffuse bands, observed towards two targets, using the CFHT/Gecko
instrument. The effect is exactly like in Figure 2 which confirms the earlier result.}
\label{Gecko}
\end{figure}

It is an important question whether the observed effects of DIB
broadening are real or, perhaps, instrumental. Fortunately,  we have in our database spectra of
the same targets, acquired using different instruments. We
compared the major DIBs: 5780 and 5797 in the spectra of HD~37061 and
HD~24912, acquired using the currently decommissioned Gecko
spectrograph, fed with the 3.6m CFH telescope at Hawaii. As
presented on Fig. \ref{Gecko} the broadening of the 5780 DIB
towards HD~37061 is real, independent of the spectrograph used to
acquire the spectrum. Let us emphasize that Gecko, being non--echelle, does not allow
to shift it's spectra to the rest wavelength velocity frame, using
some identified atomic or molecular interstellar features. Thus
Fig. \ref{Gecko} compares only profile widths and shapes.

\begin{figure}
\includegraphics[width=9cm]{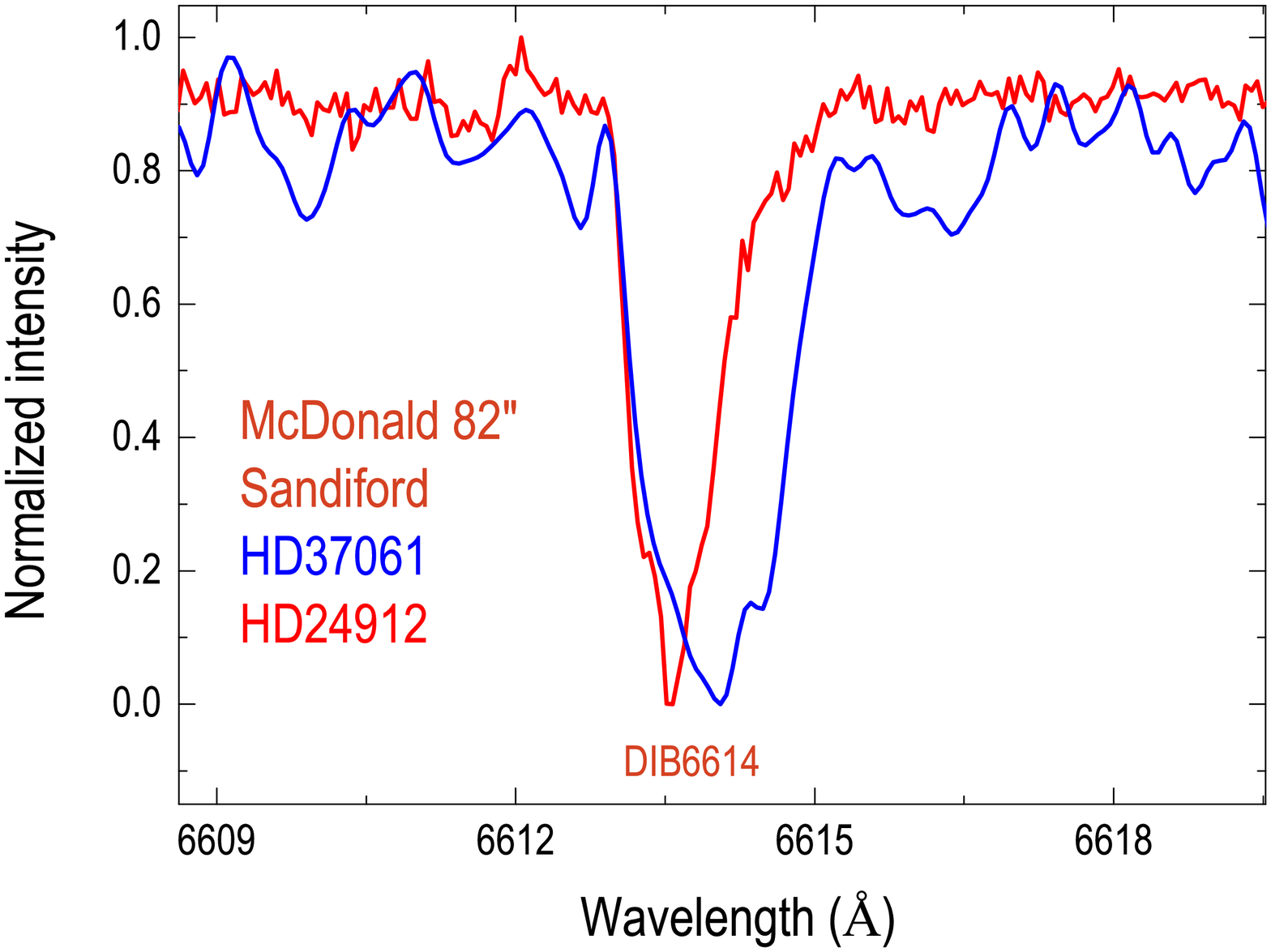}
\caption{Spectra of two of the objects, observed using the
McDonald Sandiford echelle spectrograph. The effect for the 6614 DIB is
the same as that for the major 5780 DIB.}
\label{fig8}
\end{figure}

We made a similar check for the 6614 DIB. This time we used spectra from
the Sandiford echelle spectrograph fed with the 82" McDonald Observatory telescope.
We compared a spectrum of HD~37061 with that of HD~144217 ($\sigma$ cloud standard).
Fig. \ref{fig8} proves that the excessive widths of DIBs in HD~37061 are real,
independent of the instrument used. The Sandiford echelle does not cover the whole
spectral range, available to ground-based observations. In our spectrum only the Na{\sc i}
doublet is available. Due to its saturation the interstellar rest wavelengths scale cannot be established.
Therefore Fig. \ref{fig8} cannot be used for the estimation of peak wavelength variation of diffuse bands $\lambda_{cen}$
but only for comparison of width and shape of DIBs.

\section{Conclusions}

\begin{figure}
    \centering
		\includegraphics[width=9cm]{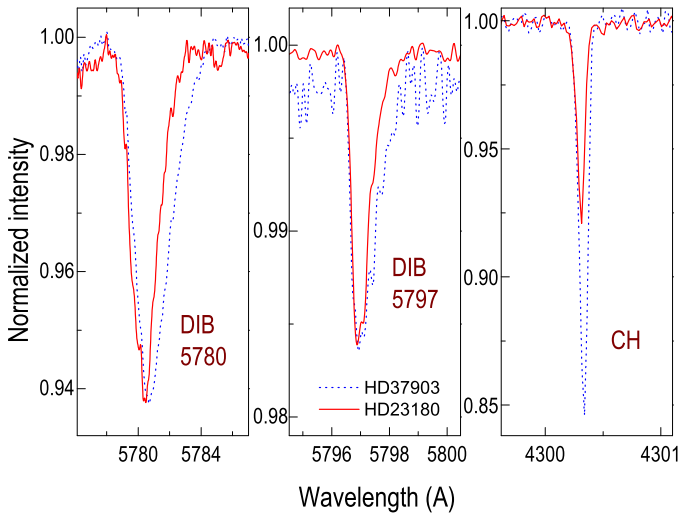}
    \caption{Broadened 5797 and 5780 DIBs profiles in HD 37903 compared with "normal" profiles from HD 23180.
    Note the lack of Doppler-split in profile of interstellar CH molecule in both objects. }
    \label{noDoppler}
 \end{figure}

The 5780, 5797, 6196, 6203 and 6614 DIBs are formed along the lines of sight where vibrationally excited levels of H$_2$ are abundantly populated which may lead to extended red wings. The broad DIBs (e.g. 5780, 6284) are also
shifted to the red. The observed DIBs broadening cannot be explained by unresolved blend of many components originated in different clouds (Fig. \ref{noDoppler});

We emphasize that the lack of correlation between FWHM and central wavelength for DIB 6196 (Fig. 2) does not imply that a broadening
effect of this DIB does not exist.  For example the extended red wing of DIB 6196 does in most cases not reach the 50\% depth
(Fig. 4). Therefore the midpoint of the FWHM of that band is almost stationary and so is the derived central wavelength despite of the
evident broadening.

What really causes the broadening and red-shift of DIB profiles? The answer can not be given before comprehensive identification of the carriers of diffuse bands.
There are several possible reasons causing the broadening of diffuse bands (mostly extending the red wing of the profile):
\begin{itemize}
  \item{
  First of all, the observed profile variations of the DIBs, cannot be caused by
thermal Doppler broadening because the observed broadening is not symmetric. Only the long wavelength side of the profile is extended. Also the observed broadening cannot be explained by Doppler splitting (Fig. \ref{noDoppler});}
  \item{
  The widths (FWHMs) of the 5780, 6196, 6203 and 6614 DIBs are broadened in clouds where the vibrationally excited levels of molecular hydrogen are populated.
The widths of these DIBs are larger for directions with larger T$^{para}_{\nu=2}$ temperatures.
However, the hydrogen temperature T$_{01}$ does not correlate with DIBs' broadening while kinetic temperature of C$_2$ does \citep{K2010}.
Indeed, DIB profiles may get broad, influenced by the rotational temperatures of homonuclear molecules, such as C$_2$
 as it was shown by \citep{K2009} and  \citep{K2010} for  6196 and 5797 \AA\ DIB and the gas kinetic and rotational temperatures of C$_2$ molecule.
 Unfortunately the bands of C$_3$ molecule are usually below the level of detection in our spectra and thus the possible correlation between kinetic and rotational
 temperatures of C$_3$ and the width of diffuse bands cannot be studied with the available data.
 Oka et al. (2010) offered to consider a possible reason for broadening and red shift in terms of rotational excitation of relatively small polar molecules.
 Marshall, Kre{\l}owski  \&  Sarre (2015) considered variations of DIB 6614 and offered an interpretation founded in terms of the population distribution among the
rotational, vibrational and potentially low-lying electronic states of a medium-sized (N$_C$ $\sim$ 20) planar PAH-type molecule.
Profile may depend on the presence of low-lying vibrational/vibronic states of the carrier molecule, i.e. on whether these transitions
are significantly wavelength-shifted from the origin band.}
 \item{
 H$_2$ vibrationally excited levels are populated by fluorescence. The broadenings of DIBs may also be caused by fluorescently populated levels.
An additional broader discussion was given in  \citet{KGMMC15} where the origin of the red shift,
observed in two Orion B1 association stars HD37020 and HD37022 was discussed. Both targets are included in the current sample;
}
\item{Another interesting  idea is  a connection between the behavior of profiles of DIBs and shapes of extinction curves.
Red-shift and broadening in molecular electronic bands might be  caused by attachment of molecules to a solid substrate (e.g. Tielens et al, 1987).
In an environment, rich of dust particles, possibly some DIB carriers can also partially attach to grain surfaces, producing what we observe in broadened DIB profiles and in shapes of extinction curves.}
\end{itemize}

Despite of the lack of decisive explanation of the red-shift phenomena, the observed broadening of diffuse bands can be used as an
additional criterion for grouping these features into families, presumably formed by the same molecule.
The use for the same purposes of red- or blue-shifts, observed in some diffuse bands, is not so straightforward because the atomic (molecular) lines
serving as a reference points for the interstellar wavelength frame and, the carriers of diffuse bands may be occupying different parts of an interstellar cloud.

\section*{Acknowledgements}
JK, RH and WS acknowledge the financial support of the Polish National Science Centre \textemdash the grant UMO-2017/25/B/ST9/01524 for the
period 2018 \textendash\ 2021. GAG acknowledges the financial support of the Ministry of Science and Higher Education of the Russian
Federation under the grant 075-15-2020-780 (N13.1902.21.0039)

This research has made use of the services of the ESO Science Archive Facility and the SIMBAD database, operated at CDS, Strasbourg, France \citep{Wanger}.
The H$_2$ temperatures were derived from observations made with the NASA/ESA Hubble Space Telescope, and obtained from the Hubble Legacy Archive,
which is a collaboration between the Space Telescope Science Institute (STScI/NASA), the Space Telescope European Coordinating Facility (ST-ECF/ESAC/ESA)
and the Canadian Astronomy Data Centre (CADC/NRC/CSA).

\section*{Data Availability}

All data used in this paper were carried out by authors and/or obtained from data archives.


\begin{thebibliography}{99}
\bibitem[Bailer-Jones et~al.(2018)] {BRFMA18} Bailer-Jones, C. A. L., Rybizki, J., Fouesneau, M., et al., 2018, AJ, 156, 58
\bibitem[Bernstein  et~al.(2003)] {Ber03} Bernstein, R., Shectman, S. A., Gunnels, et al., 2003, SPIE, 4841, 1694
\bibitem[Campbell et al.(2015)]{2015Natur.523..322C} Campbell, E.~K., Holz, M., Gerlich, D., et al.\ 2015, \nat, 523, 322
\bibitem[Dahlstrom  et~al. (2013)] {Dahlstrom2013} Dahlstrom, J., York, D.G., Welty, D.E.,  et~al., 2013, ApJ,  773, 41
\bibitem[Dekker  et~al. (2000)]{Dekal00} Dekker, H., D'Odorico, S., Kaufer, A.,  et~al., 2000, SPIE,  4008, 534
\bibitem[Fan  et~al. (2019)]{Fan2019} Fan, H., Hobbs, L. M., Dahlstrom, J. A., et~al., 2019, ApJ,  878, 151
\bibitem[Galazutdinov  et~al. (2006)]{G06} Galazutdinov, G. A., Manic\'{o}, G., \& Kre{\l}owski, J., 2006, MNRAS,  366, 1075 
\bibitem[Galazutdinov, LoCurto \& Kre{\l}owski (2008)]{GLK08} Galazutdinov, G. A., LoCurto, G. \& Kre{\l}owski, J., 2008, ApJ,  682, 1076
\bibitem[Galazutdinov et al.(2008)]{2008PASP..120..178G} Galazutdinov, G.~A., LoCurto, G., Han, I., et al.\ 2008, \pasp, 120, 178  
\bibitem[\protect\citeauthoryear{Galazutdinov et al.}{2017}]{2017MNRAS.465.3956G} Galazutdinov G.~A., Shimansky V.~V., Bondar A., Valyavin G., Kre{\l}owski J., 2017, MNRAS, 465, 3956. doi:10.1093/mnras/stw2948
\bibitem[\protect\citeauthoryear{Galazutdinov \& Kre{\l}owski}{2017}]{2017AcA....67..159G} Galazutdinov G.~A., Kre{\l}owski J., 2017, AcA, 67, 159. doi:10.32023/0001-5237/67.2.4
\bibitem[Galazutdinov et al.(2020)]{2020AJ....159..113G} Galazutdinov, G., Bondar, A., Lee, B.-C., et al., 2020, \aj, 159, 113
\bibitem[\protect\citeauthoryear{Galazutdinov et al.}{2021}]{2021AJ....161..127G} Galazutdinov G.~A., Valyavin G., Ikhsanov N.~R., Kre{\l}owski J., 2021, AJ, 161, 127. doi:10.3847/1538-3881/abd4e5
\bibitem[Gnaci{\'n}ski (2009)]{Gnacinski2009} Gnaci{\'n}ski P., 2009, AcA,  59, 325
\bibitem[Gnaci{\'n}ski (2011)]{Gnacinski2011} Gnaci{\'n}ski P., 2011, A\&A,  532, A122
\bibitem[Gnaci{\'n}ski (2013)]{Gnacinski2013} Gnaci{\'n}ski P., 2013, A\&A,  549, A37
\bibitem[Gnaci{\'n}ski et al. (2016)]{Gnacinski2016} Gnaci{\'n}ski P., Weselak T.; Krelowski J., 2016, AcA, 66, 121
\bibitem[Herbig \& Soderblom (1982)]{HS82} Herbig, G. H. \& Soderblom, D. R., 1982, ApJ,  252, 610
\bibitem[Jensen  et~al. (2010)]{Jensen} Jensen A.G., Snow T.P., Sonneborn G., Rachford B.L., 2010, ApJ,  711, 1236
\bibitem[Josafatsson \& Snow (1987)]{JS87} Josafatsson, K. \& Snow, T. P., 1987, ApJ,  319, 436
\bibitem[Kaufer  et~al. (1999)]{Ketal99} Kaufer, A., Stahl, O., Tubbesing, S.,  et~al., 1999, The Messenger,  95, 8
\bibitem[Ka{\'z}mierczak  et~al. (2009)]{K2009} Ka\'zmierczak, M., Gnaci\'{n}ski, P., Schmidt, M.R., et~al., 2009, A\&A,  498,  785
\bibitem[Ka{\'z}mierczak et al.(2010)]{K2010} Ka{\'z}mierczak, M., Schmidt, M.~R., Galazutdinov, G.~A., et al.\ 2010, \mnras, 408, 1590
\bibitem[Kerr  et~al. (1998)]{Kerr1998} Kerr, T.H.,  Hibbins, R.E., Fossey, S.J.,   et~al., 1998, ApJ,  495, 941
\bibitem[Kim  et~al. (2007)]{kimetal2007} Kim, K.-M., Han, I., Valyavin, G. G.,  et~al., 2007, PASP,  119, 1052
\bibitem[Kre{\l}owski \& Walker (1987)]{KG87} Kre{\l}owski, J. \& Walker, G.A.H., 1987, ApJ,  312, 860
\bibitem[Kre{\l}owski \& Westerlund (1988)]{KW88} Kre{\l}owski, J. \& Westerlund, B.E., 1988, A\&A,  190, 339
\bibitem[Kre{\l}owski \& Sneden (1995)]{KS95} Kre{\l}owski, J. \& Sneden, C., 1995, in Tielens \& Snow (eds.) ``The Diffuse Interstellar Bands'', Kluwer, p. 13
\bibitem[Kre{\l}owski \& Greenberg (1999)]{KG99} Kre{\l}owski, J. \& Greenberg, J.M., 1999, A\&A,  346, 199
\bibitem[Kre{\l}owski  et~al. (2011)]{KGB11} Kre{\l}owski, J., Galazutdinov, G. \&  Beletsky, Y., 2011, A\&A,  531, 68
\bibitem[Kre{\l}owski  et~al. (2015)]{KGMMC15} Kre{\l}owski, J., Galazutdinov, G. A., Mulas, G. et al., 2015, MNRAS,  451, 3210
\bibitem[Kre{\l}owski (2018)]{JK18} Kre{\l}owski, J., 2018, PASP,  130, 1001
\bibitem[Kre{\l}owski et al. (2019a)]{KSGB19} Kre{\l}owski, J., Strobel, A., Galazutdinov, G.A., et al., 2019, MNRAS, 486, 112
\bibitem[Kre{\l}owski et al.(2019b)]{2019AcA....69..369K} Kre{\l}owski, J., Mari{\'c}, T., Karipis, A., et al.\ 2019, AcA, 69, 369 
\bibitem[Marshall, Kre{\l}owski  \&  Sarre (2015)]{MKS19} Marshall C.C.M., Krelowski J. and Sarre P.J., 2015, MNRAS, 453, 3912
\bibitem[Mayor et al. (2003)]{May03} Mayor, M., Pepe, F., Queloz, D., et al., 2003, The Messenger 114, 20
\bibitem[McCarthy et al. (1993)]{MSBB93} McCarthy, J. K., Sandiford, B. A., Boyd, et al., 1993, PASP 105, 881
\bibitem[Megier et al.(2005)]{2005ApJ...634..451M} Megier, A., Strobel, A., Bondar, A., et~al., 2005, \apj, 634, 451
\bibitem[Megier et al.(2009)]{2009A&A...507..833M} Megier, A., Strobel, A., Galazutdinov, G.~A., et~al., 2009, \aap, 484, 381
\bibitem[Motylewski  et~al. (2000)]{Motylewski2000} Motylewski, T., Linnartz, H., Vaizert, O.,  et~al., 2000, ApJ,  531, 312
\bibitem[Oka et~al. (2013)]{OWJ13} Oka, T., Welty, D.E., Johnson, S., et al., 2013, ApJ, 773, 42
\bibitem[Salama  et~al. (1999)]{Salama1999} Salama,  F., Galazutdinov, G.A., Kre{\l}owski, J., et~al., 1999, ApJ,  526, 265
\bibitem[Salama (2008)]{Salama2008} Salama F., 2008, IAU Symp.,  251, 357
\bibitem[Sarre  et~al. (1995)]{Sarre1995} Sarre, P., Miles, J.R., Kerr, T.H., et~al., 1995, MNRAS,  277, L41
\bibitem[Schmidt et~al. (2014)]{Schmidt2014} Schmidt, M., Kre{\l}owski, J., Galazutdinov, G.A., et~al., 2014, MNRAS,  441, 1134
\bibitem[Siebenmorgen et~al. (2020)]{Siebenmorgen2020} Siebenmorgen R., Krelowski J., Smoker J., et al. 2020, A\&A, 641, 35
\bibitem[\protect\citeauthoryear{Spitzer, Dressler, \& Upson}{1964}]{1964PASP...76..387S} Spitzer L., Dressler K., Upson W.~L., 1964, PASP, 76, 387. doi:10.1086/128122
\bibitem[Tielens et~al. (1987)]{Tielens1987} Tielens A. G. G. M., Allamandola L. J., Hollenbach D. J., et~al., 1987, Astrophysics and Space Science Library, Vol. 134, Interstellar Processes, Berlin Springer-Verlag, 397
\bibitem[Wanger (2000)]{Wanger} Wanger, M., Ochsenbein, F., Egret, D., et~al., 2000, A\&AS,  143, 9
\bibitem[Zhao  et~al. (2015)]{Zhao2015} Zhao, D., Galazutdinov, G.A., Linnartz, H., et~al., 2015, ApJ, 805, 12
\end{thebibliography}
\end{document}